\pdfoutput=1
\documentclass[11pt]{article}

\usepackage[margin=1in]{geometry}
\usepackage{amsmath,amssymb}
\usepackage{tikz-cd}
\usepackage{multicol}
\usepackage{authblk}
\usepackage{setspace}
\usepackage[hidelinks]{hyperref}

\title{Geometric Buoyancy-like Effects of Static Structures with Internal Stress in Schwarzschild Spacetime}
\author{Yuji Takeuchi}
\affil{Tomonaga Center for the History of the Universe, Institute of Pure and Applied Sciences, University of Tsukuba, Tennodai 1-1-1, Tsukuba, Ibaraki 305-8571, Japan}
\date{}
\newlength{\keywordlabelwidth}
\newcommand{\keywords}[1]{%
  \par\noindent
  \textbf{Keywords: }%
  \parbox[t]{\dimexpr\textwidth-\keywordlabelwidth\relax}{\raggedright #1}%
  \par
}

\begin{document}


\maketitle

\begin{abstract} 
In curved spacetime, deformable bodies can undergo a displacement of their center of mass through cyclic internal motions without the use of propellant, as shown by Wisdom. In this paper, we explicitly construct static structures with internal stress in Schwarzschild spacetime that generate a buoyancy-like force (in the sense of a pressure-imbalance-induced force, but with a different physical origin from fluid buoyancy) purely from stress distribution, without non-gravitational external forces or cyclic internal motions. The mechanism is illustrated using simple static structures composed of rod-like elements aligned along spatial geodesics with assigned internal stresses, for which both numerical calculations and perturbative analyses are performed. The resulting effect is extremely small for realistically sized structures and does not lead to actual ascent against gravity. Nevertheless, it reveals a new aspect of extended-body dynamics in curved spacetimes, namely how stress-curvature coupling can influence motion of extended bodies even in static configurations.
\end{abstract}


\begin{center}
\textit{E-mail:} \href{mailto:takeuchi@hep.px.tsukuba.ac.jp}{takeuchi@hep.px.tsukuba.ac.jp}
\end{center}

\section{Introduction}

Wisdom proposed the so-called curved-spacetime swimming effect~\cite{wisdom}, in which a deformable body can achieve a displacement of its center of mass through cyclic internal motions, without the use of any propellant, in a curved spacetime. This effect was initially discussed based on intuitive mechanical analogies, but subsequent studies have reinterpreted it within a more general framework of extended-body dynamics. Indeed, Silva et al.~\cite{silva} pointed out that Wisdom's original analysis contains certain formal issues; nevertheless, using a covariant formulation based on Dixon~\cite{dixon}, they showed that the swimming effect can be understood as arising from the coupling between spacetime curvature and multipole moments of quadrupole order and higher.

In Dixon's multipole formalism, multipole moments of quadrupole order and higher, arising from internal stresses, are known to couple to curvature gradients and thereby contribute to the motion of the center of mass as a deviation from geodesic motion. However, to the best of the author's knowledge, no explicit and concrete model has been constructed in which such effects arise in a static structure without time-dependent deformations. In this work, we explicitly construct a static structure in Schwarzschild spacetime composed of rod-like elements arranged along geodesics and carrying stress only in their tangential directions (hereafter referred to as geodesic rods), and show that an effective buoyancy-like force can arise without the presence of any non-gravitational external forces.

While such curvature-induced effects have been discussed primarily in the context of time-dependent internal motions, it remains less explored whether similar effects can arise in purely static configurations. In particular, it is not obvious whether internal stresses alone, without any time dependence, can lead to a net effective force in a curved spacetime. Addressing this requires an explicit construction of extended structures in which the interplay between geometry and stress can be examined concretely.

To elucidate the underlying mechanism, we explicitly construct static structures composed of geodesic rods as their fundamental building elements. Each geodesic rod, when considered in isolation, has equal tension or compression at its two endpoints and does not exhibit any imbalance of internal stress. However, in Schwarzschild spacetime, where the gravitational field depends on position, the situation differs fundamentally when such rods are combined to form a structure.
For example, consider a system composed of multiple vertices connected by geodesic rods. Even if the configuration is chosen such that the stresses at the rod endpoints balance at a given vertex, it does not necessarily follow that the balance holds simultaneously at other vertices. This is because, in Schwarzschild spacetime, the tangential directions of geodesics depend on the radial position, so that the spatial directions of the stresses along the rods do not coincide from one vertex to another. In other words, even if the magnitudes of the stresses are matched, their directions fail to align globally due to spacetime curvature, thereby leaving a residual imbalance.
As a result, even though the internal stresses in each individual geodesic rod are completely balanced, a mismatch in their directions remains when the system is considered as a whole, and an effective force can arise at the level of the entire structure. The buoyancy-like force demonstrated in this work can be interpreted as originating from such an interplay between internal stresses and curvature gradients, and is distinguished by the fact that it requires neither non-gravitational external forces nor time-dependent internal motions.

\section{Geodesic rods in curved spacetime}
\label{sec:geodesic-rods}

\subsection{Geodesic rods}

We define a \textit{geodesic rod} as a one-dimensional static structural element arranged along a spatial geodesic lying on a constant-time hypersurface $t=\mathrm{const.}$ in the static coordinate system of Schwarzschild spacetime. The centerline of the rod coincides with a spatial geodesic, and both its shape and internal state are assumed to be time-independent.

The rod supports internal stress only in the form of tension or compression along the tangential direction of its centerline, while bending moments and shear stresses are neglected. At each point along the rod, the stress acts along the tangential direction of the centerline, and its orientation is described with respect to a local orthonormal frame (tetrad).

At the endpoints of the rod, the stresses act in opposite directions along the tangential direction of the centerline. However, their magnitudes are not constant along the rod; they vary with position due to the gravitational field. Nevertheless, when evaluated by static observers defined with respect to infinity, the stress is rescaled by the lapse function---relating coordinate time to local proper time---and becomes constant along the rod. As a result, for an isolated geodesic rod, the endpoint stresses balance exactly, and no net imbalance of internal stress arises.

Details of the treatment of stress transmission along the rod, including both the massless case and the case with finite rod mass, are provided in Appendix~\ref{app:stress-propagation}, based on a covariant formulation of extended-body dynamics. Furthermore, although the geodesic rod structures considered in this work contain internal tension or compression that can act as a source of gravity through the energy--momentum tensor, their self-gravity is neglected in the present analysis. That is, the structure is treated as a test structure in a fixed Schwarzschild spacetime.

\subsection{Spatial geodesics}

To describe the geometry along which geodesic rods are arranged, we consider spatial geodesics on constant-time hypersurfaces in Schwarzschild spacetime. The Schwarzschild metric is given by
\begin{equation}
ds^2 = -\alpha^2(r)\,dt^2 + \alpha^{-2}(r)\,dr^2 + r^2\left(d\theta^2 + \sin^2\theta\,d\varphi^2\right),
\label{eq:schwarzschild_metric}
\end{equation}
where $\alpha(r) \equiv \sqrt{1-r_s/r}$ is the lapse function and $r_s$ is the Schwarzschild radius. For simplicity, we restrict the configuration to the equatorial plane $\theta = \pi/2$. To describe static structures, we consider the spatial metric on a constant-time hypersurface $t=\mathrm{const.}$,
\begin{equation}
dl^2 = \alpha^{-2}(r)\,dr^2 + r^2\,d\varphi^2.
\label{eq:spatial_metric}
\end{equation}

The centerline of a geodesic rod coincides with a spatial geodesic of the above metric (see the previous subsection for the definition). Therefore, its shape is determined by the geodesic equation associated with the spatial metric in Eq.~\eqref{eq:spatial_metric}. We parametrize the spatial geodesic by $\varphi$, and introduce
\begin{equation}
u(\varphi) \equiv \frac{1}{r(\varphi)},
\end{equation}
\begin{equation}
v(\varphi) \equiv \frac{du}{d\varphi}.
\end{equation}
Introducing a conserved quantity $J$ along the geodesic, we obtain
\begin{equation}
\frac{1}{J^2} = u^2 + \frac{v^2}{\alpha^2},
\qquad
\alpha^2 = 1-r_s u.
\end{equation}
Using this conserved quantity, the spatial geodesic $u(\varphi)$ is determined by the ordinary differential equation
\begin{equation}
\frac{d^2 u}{d\varphi^2} + u = -\frac{r_s}{2J^2} + \frac{3}{2} r_s u^2,
\label{eq:spatial_geodesic_ode}
\end{equation}
together with the initial conditions $u(\varphi_0)$ and $v(\varphi_0)$ at a given endpoint $\varphi_0$.

An orthonormal frame (tetrad) $\{e_{(\hat r)}, e_{(\hat \varphi)}\}$ associated with the spatial metric in Eq.~\eqref{eq:spatial_metric} is given by
\begin{equation}
e_{(\hat r)} = \alpha\,\partial_r,
\qquad
e_{(\hat \varphi)} = \frac{1}{r}\,\partial_\varphi.
\end{equation}
The (unnormalized) tangent vector to the curve $u(\varphi)$ is
\begin{equation}
\frac{d}{d\varphi}
= \frac{dr}{d\varphi}\,\partial_r + \partial_\varphi
= \alpha^{-1} u^{-2}\left(-v\,e_{(\hat r)} + \alpha u\,e_{(\hat \varphi)}\right).
\end{equation}
Accordingly, the unit tangent vector $\hat t$, and the unit normal vector $\hat n$ defined counterclockwise with respect to $\{e_{(\hat r)}, e_{(\hat \varphi)}\}$, are given by
\begin{equation}
\hat t = \frac{-v\,e_{(\hat r)} + \alpha u\,e_{(\hat \varphi)}}{\sqrt{v^2 + \alpha^2 u^2}},
\qquad
\hat n = \frac{-\alpha u\,e_{(\hat r)} - v\,e_{(\hat \varphi)}}{\sqrt{v^2 + \alpha^2 u^2}}.
\end{equation}

These geometrical quantities provide the basis for describing the direction of stress and its propagation along geodesic rods. In the next section, we analyze the force balance and the resulting imbalance of stresses in such geodesic rod structures.

\section{Four-vertex geodesic-rod structure: diamond configuration}

\subsection{Configuration of the diamond configuration}

\begin{figure}[tb]
\centering
\includegraphics[width=0.6\linewidth]{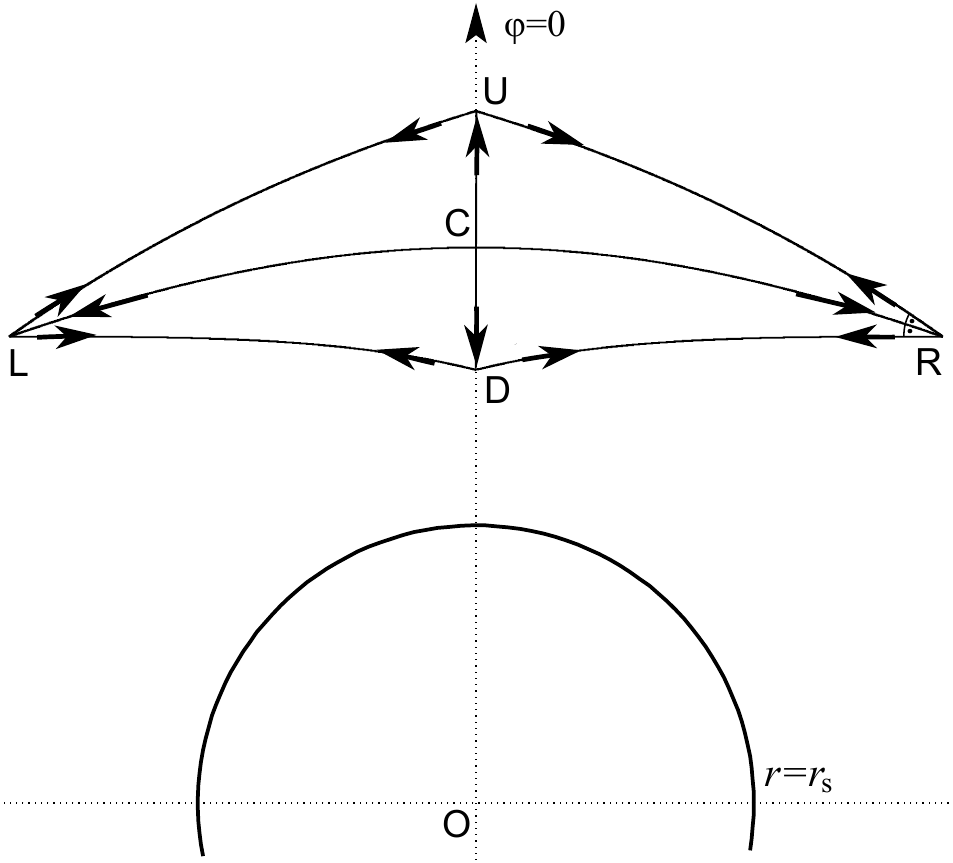}
\caption{An example of the diamond configuration on the equatorial plane ($\theta=\pi/2$) of Schwarzschild spacetime, shown in the $(r,\varphi)$ plane (configuration obtained by numerical calculation; see text). The vertices $U$, $D$, $L$, $R$, the midpoint $C$ of the rod $LR$, and the origin $O$ are indicated. Arrows represent the directions of forces associated with the stresses along the geodesic rods. The angles $\angle CRU$ and $\angle CRD$ are equal and correspond to $\eta$. The circle denotes $r=r_s$ (the Schwarzschild radius).}
\label{fig:diamond-configuration}
\end{figure}

As shown in Fig.~\ref{fig:diamond-configuration}, we consider a static structure composed of four vertices $U$, $D$, $L$, and $R$ connected by geodesic rods. The rods connecting the vertices are $LR$, $UD$, $LU$, $RU$, $LD$, and $RD$, and all are arranged along spatial geodesics defined in Sec.~\ref{sec:geodesic-rods}.
Let $C$ denote the midpoint of the rod $LR$, which is located at $(r,\varphi)=(r_0,0)$. The vertices $U$, $D$, and the point $C$ lie on $\varphi=0$, and the entire structure is symmetric with respect to $\varphi=0$. The rods $RU$ and $RD$ are assumed to form an angle $\eta$ with the rod $LR$ at the point $R$. Here, the angle $\eta$ is measured with respect to the local orthonormal frame (tetrad) defined in Sec.~\ref{sec:geodesic-rods}.

This structure (hereafter referred to as the \textit{diamond configuration}) is uniquely specified by the geometric parameters $(r_0,\varphi_0,\eta)$.
First, we place the midpoint $C$ of the rod $LR$ at $(r_0,0)$, and solve the geodesic equation Eq.~\eqref{eq:spatial_geodesic_ode} with the initial condition $(u_c,v_c)=(r_0^{-1},0)$ up to $\varphi=\varphi_0$ to obtain the point $L$. By symmetry, the point $R$ is determined at $\varphi=-\varphi_0$.
Next, we evaluate the unit tangent vector $\hat t$ and the unit normal vector $\hat n$ of the rod $LR$ at the point $R$, and construct the geodesics $RU$ and $RD$ by taking initial directions that form an angle $\eta$ with these vectors. By integrating each geodesic up to $\varphi=0$, we obtain the vertices $U$ and $D$.
Through this procedure, the diamond configuration consisting of the vertices $R$, $L$, $U$, $D$ and the six geodesic rods is uniquely determined for a given set of parameters $(r_0,\varphi_0,\eta)$.

Finally, we assume the following configuration of internal stresses. The rods $LR$ and $UD$ are under compression, while the rods $LU$, $RU$, $LD$, and $RD$ are under tension with equal magnitude $S$. Here, $S$ denotes the tension evaluated by static observers defined with respect to infinity (see Appendix~\ref{app:stress-propagation}). This stress configuration is consistent with the case in which the entire structure remains static in flat spacetime.

\subsection{Force balance at vertices}

We first consider the force balance at the vertex $R$. We assume that the compressive force from the rod $LR$ is balanced by the tensions from the rods $RU$ and $RD$. As defined in Sec.~\ref{sec:geodesic-rods}, the forces at the endpoints of each rod act along the tangential direction of the rod. Under this assumption, the force balance at $R$ can be written as
\begin{equation}
f_{RU}^{(R)} = f_{RD}^{(R)},
\qquad
f_{LR}^{(R)} = \left(f_{RU}^{(R)} + f_{RD}^{(R)}\right) \cos\eta,
\end{equation}
where the superscript indicates the vertex at which the force is evaluated, and the subscript denotes the geodesic rod that transmits the force. It is thus possible to construct the configuration such that the endpoint forces of the geodesic rods are exactly balanced at the vertex $R$, and similarly at the vertex $L$.

We next examine the force balance at the vertices $U$ and $D$. At the vertex $U$, the tensions along the rods $RU$ and $LU$ cancel in the $e_{(\hat \varphi)}$ component, while the sum of their $e_{(\hat r)}$ components provides the compressive force transmitted to the rod $UD$. A similar relation holds at the vertex $D$, where the sum of the $e_{(\hat r)}$ components of the tensions along $RD$ and $LD$ contributes to the compression of the rod $UD$.
In flat spacetime, the tangential directions of the rods $RU$, $LU$ and $RD$, $LD$ are symmetric with respect to the horizontal axis, and hence the magnitudes of these compressive contributions at $U$ and $D$ are equal. As a result, no imbalance of internal stress arises along the rod $UD$. In contrast, in Schwarzschild spacetime, the tangential directions at $U$ and $D$ are no longer symmetric. Consequently, a nonvanishing internal stress imbalance appears along the rod $UD$. To maintain a static configuration, this imbalance must be compensated by applying an appropriate external force (external load) at either $U$ or $D$.

To quantify this imbalance, we define the unit tangent vector of the rod $RU$ at the vertex $U$ to point in the direction from $R$ to $U$, and denote its $e_{(\hat r)}$ component by $t_r^{(U)}$. Similarly, we define $t_r^{(D)}$ for the rod $RD$ at the vertex $D$. Using the tension magnitude $S$ acting along the rods $RU$ and $RD$, and taking into account the contributions from the two rods on each side, the radial component of the resulting imbalance of internal stress along the rod $UD$ is given by
\begin{equation}
F_r = -2S\left(t_r^{(U)} + t_r^{(D)}\right).
\label{eq:diamond_force}
\end{equation}
Here, the negative sign reflects the fact that the tension $S$ acts so as to pull the endpoints $U$ and $D$ inward.

\subsection{Numerical evaluation}

We numerically evaluate the radial force $F_r$ acting on the diamond configuration composed of geodesic rods defined in the previous subsection. As a representative example in which the effect can be clearly demonstrated, we consider a spacetime with $r_s=0.1$ and choose $(r_0,\varphi_0,\eta)=(0.2,45^\circ,20^\circ)$. Although this setup is physically unrealistic, it is useful as an explicit example demonstrating that the effect discussed in this work is indeed nonzero.

For a given set of parameters $(r_0,\varphi_0,\eta)$, we numerically construct the diamond configuration by solving for the geodesics that define each rod. The geodesic equation is integrated using the fourth-order Runge--Kutta method (RK4). From the resulting geodesics, we evaluate the tangent vectors at each vertex and compute $F_r/S$ using Eq.~\eqref{eq:diamond_force}. We obtain
\begin{equation}
\frac{F_r}{S} = 0.030.
\label{eq:diamond_numeric_force}
\end{equation}
We have verified that the discretization error in the numerical integration is sufficiently small compared to this value, indicating that a significant positive radial force arises in the diamond configuration. The configuration shown in Fig.~\ref{fig:diamond-configuration} is drawn based on this numerical result.

\subsection{Perturbative analysis}

In this subsection, we analyze the nonvanishing force demonstrated in the numerical calculation within the regime of a weak gravitational field and small angular scale of the structure, $r_0 \gg r_s$ and $\varphi_0 \ll 1$.
We first solve the spatial geodesic equation perturbatively by expnading $u(\varphi)$ in powers of $r_s$ as
\begin{equation}
u(\varphi) = u_0(\varphi) + r_s u_1(\varphi) + \cdots.
\end{equation}
Substituting this expansion into Eq.~\eqref{eq:spatial_geodesic_ode}, we obtain the equations at each order in $r_s$. The leading-order equation is
\begin{equation}
u_{0}'' + u_{0} = 0,
\end{equation}
whose solution is given by
\begin{equation}
u_{0}(\varphi) = u_{i} \cos(\varphi - \varphi_{0}) + v_{i} \sin(\varphi - \varphi_{0}),
\end{equation}
with $u_i \equiv  u(\varphi_0)$ and $v_i \equiv v(\varphi_0)$. At first order in $r_s$, expanding Eq.~\eqref{eq:spatial_geodesic_ode} consistently in $r_s$, including the $r_s$ dependence of the conserved quantity $J$, we obtain an inhomogeneous equation of the form
\begin{equation}
u_{1}'' + u_{1} = -\frac{1}{2}(u_{i}^2 + v_{i}^2) + \frac{3}{2}u_{0}^2,
\end{equation}
where the source term is determined by the zeroth-order solution $u_0$. The solution for $u_1$ can be obtained analytically with initial conditions $u_1(\varphi_0)=0$ and $u_1'(\varphi_0)=0$.
Using these perturbative solutions, we evaluate $t_r^{(U)}$ and $t_r^{(D)}$ for the rods $RU$ and $RD$ at the vertices $U$ and $D$, respectively, and compute $F_r/S$ up to first order in $r_s$ based on Eq.~\eqref{eq:diamond_force}. The resulting expression for $F_r/S$ can be written as a polynomial in trigonometric functions of $\varphi_0$. Expanding further in the limit $\varphi_0 \ll 1$, the leading nonvanishing term appears at first order in $r_s$ and is proportional to $\varphi_0^3$. Specifically,
\begin{equation}
  \frac{F_r}{S} = r_s\left[\frac{1}{2r_0} \sin(\eta)\tan(\eta)\,\varphi_0^3 + O(\varphi_0^5)\right] + O(r_s^2).
  \label{eq:diamond_force_pert}
\end{equation}

This result shows that $F_r$ vanishes in the flat-spacetime limit, and that the nonzero value of $F_r/S$ obtained numerically originates from spacetime curvature. In particular, the effect arises from the fact that the directions of stresses evaluated at different radial positions do not coincide due to spatial variations of the curvature. Therefore, the present result can be regarded as a simple example illustrating how the coupling between curvature and internal stresses gives rise to a nonvanishing force in a concrete structure composed of geodesic rods.
Furthermore, this perturbative result is consistent with the numerical results presented above, confirming the validity of the approximation for small $\varphi_0$ and $r_s/r_0 \ll 1$.

However, to estimate the scale of the present effect for a realistically sized structure on the Earth's surface, we take $r_0$ to be the Earth's radius and adopt an angular scale $\varphi_0$ corresponding to a length scale of $1\,\mathrm{m}$, together with $\eta = 20^\circ$. Evaluating $F_r/S$ for the diamond configuration under these conditions yields $\sim 3\times 10^{-31}$. This value is extremely small and would be practically impossible to detect experimentally.

\section{Four-vertex geodesic-rod structure: triangle configuration with an internal node}

\subsection{Configuration and numerical evaluation of the triangle and inverted triangle configurations}

In addition to the diamond configuration considered in the previous section, we examine, as another simple example, the triangle configuration and the inverted triangle configuration with an internal node.

\begin{figure}[tb]
\centering
\includegraphics[width=0.6\linewidth]{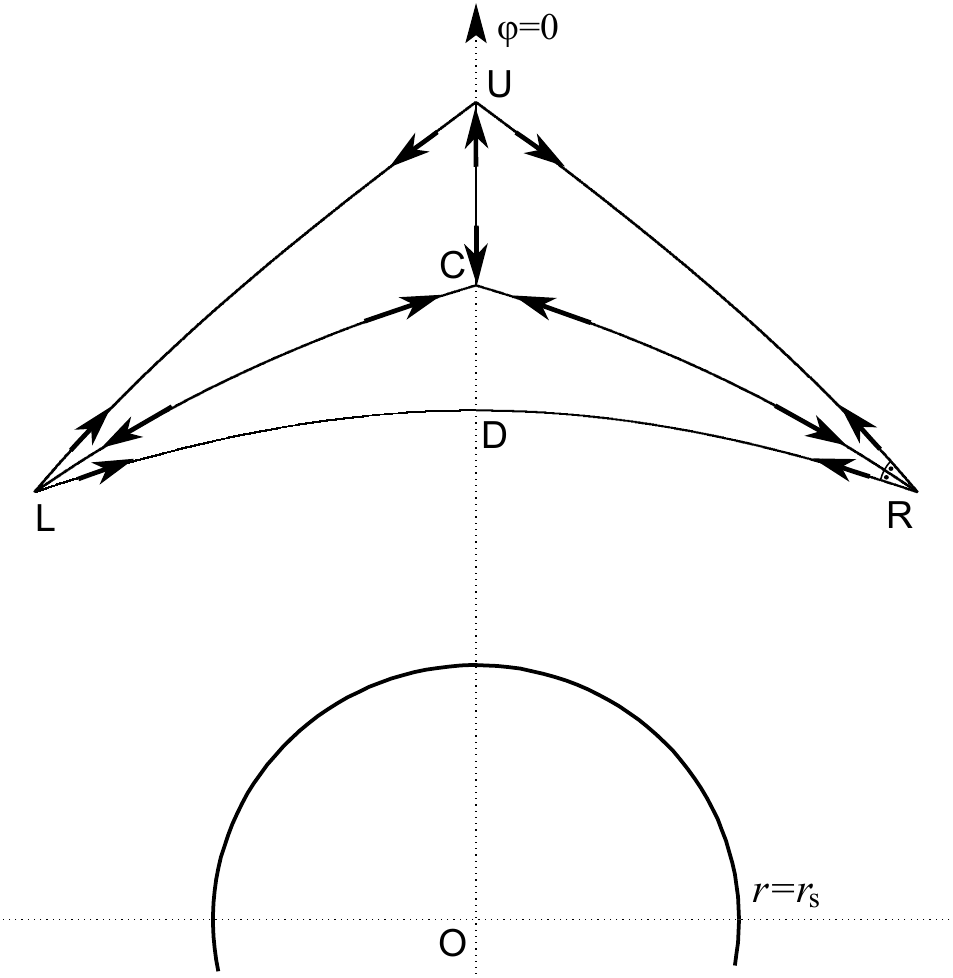}
\caption{An example of the triangle configuration (configuration obtained by numerical calculation; see text). The vertices $U$, $L$, $R$, the internal node $C$, and the midpoint $D$ of the rod $LR$ are indicated. Arrows represent the directions of forces associated with the stresses along the geodesic rods. The angles $\angle CRU$ and $\angle CRD$ are equal and correspond to $\eta$.}
\label{fig:triangle-configuration}
\end{figure}

The triangle configuration, shown in Fig.~\ref{fig:triangle-configuration}, consists of three vertices $U$, $L$, and $R$ forming an outer triangle, together with an internal node $C$. These points are connected by six geodesic rods $LR$, $LU$, $RU$, $LC$, $RC$, and $UC$. We define $D$ as the midpoint of the base $LR$. The points $U$, $C$, and $D$ all lie on $\varphi=0$, and the entire structure is symmetric with respect to $\varphi=0$.
First, we place the point $D$ at $(r,\varphi)=(r_0,0)$. From this point, we solve the spatial geodesic equation Eq.~\eqref{eq:spatial_geodesic_ode} with the initial condition $(u_d,v_d)=(r_0^{-1},0)$, and define the point at $\varphi=\varphi_0$ as $L$. By symmetry, the point $R$ is determined at $\varphi=-\varphi_0$, which fixes the base $LR$.
Next, we evaluate the unit tangent vector $\hat t$ and the unit normal vector $\hat n$ of the rod $LR$ at the point $R$, and construct geodesics by taking initial directions that form angles $\eta$ and $2\eta$ with respect to the rod $LR$. By integrating these geodesics up to $\varphi=0$, we obtain the points $C$ and $U$, respectively. By symmetry, the rods $LC$ and $LU$ are obtained similarly from the point $L$. Finally, the structure is completed by connecting the points $U$ and $C$ with a geodesic rod $UC$.
From this construction, the relations $\angle CRU = \angle CRD = \eta$ hold at the point $R$, and similarly $\angle CLU = \angle CLD = \eta$ at the point $L$. In this way, the structure is uniquely specified by the geometric parameters $(r_0,\varphi_0,\eta)$.

We assume the following configuration of internal stresses. The rods $RC$, $LC$, and $UC$ are under compression, while the rods $LR$, $RU$, and $LU$ are under tension with equal magnitude $S$. This stress configuration is consistent with the case in which the entire structure remains static in flat spacetime.
In Schwarzschild spacetime, however, even if the stresses are chosen so that compression and tension balance at the vertices $L$ and $R$, the compressive forces between the points $U$ and $C$ do not cancel, resulting in an imbalance of internal stress.

To quantify this imbalance, we numerically evaluate the radial force $F_r$ for the triangle configuration following the same procedure as for the diamond configuration. We adopt the parameters $r_s=0.1$ and $(r_0,\varphi_0,\eta)=(0.2,45^\circ,20^\circ)$. In this case, we obtain
\begin{equation}
\frac{F_r}{S} = 0.020.
\label{eq:triangle_numeric_force}
\end{equation}
This result shows that an upward effective force arises also in the triangle configuration. The configuration shown in Fig.~\ref{fig:triangle-configuration} is drawn based on this numerical result.

\begin{figure}[tb]
\centering
\includegraphics[width=0.6\linewidth]{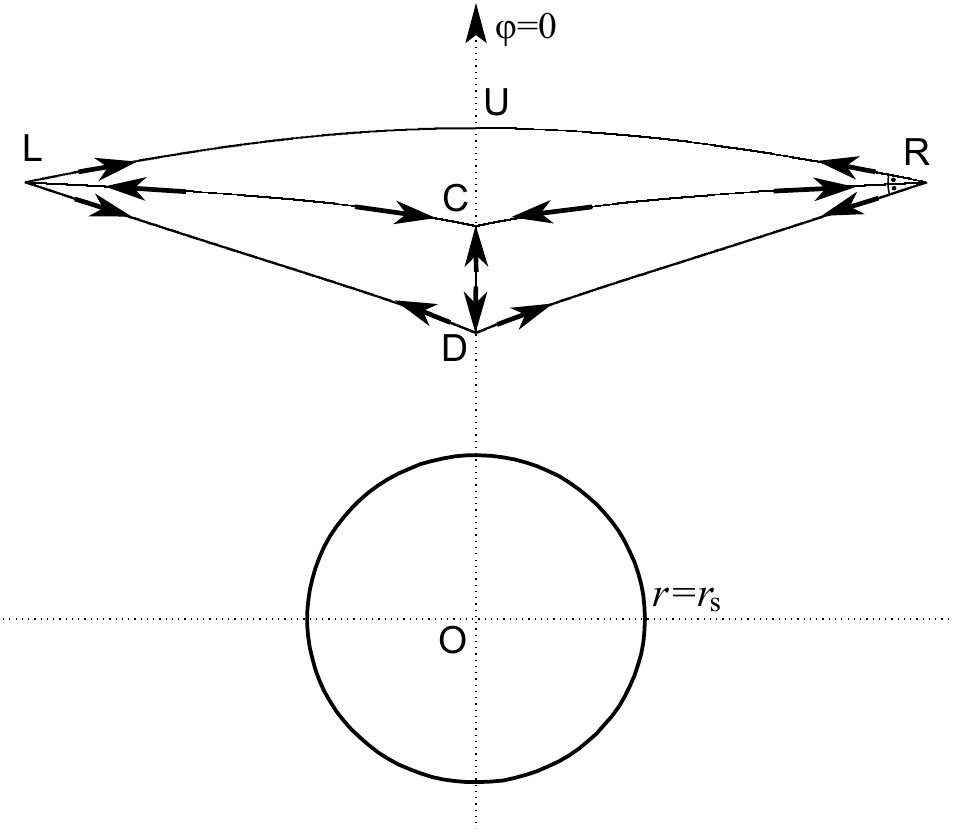}
\caption{An example of the inverted triangle configuration (configuration obtained by numerical calculation; see text). The vertices $D$, $L$, $R$, the internal node $C$, and the midpoint $U$ of the rod $LR$ are indicated. Arrows represent the directions of forces associated with the stresses along the geodesic rods. The angles $\angle CRU$ and $\angle CRD$ are equal and correspond to $\eta$.}
\label{fig:inverted-triangle-configuration}
\end{figure}

Similarly, as shown in Fig.~\ref{fig:inverted-triangle-configuration}, we consider the inverted triangle configuration obtained by exchanging the roles of the points $U$ and $D$ in the triangle configuration. The construction proceeds in the same manner, with corresponding vertices, internal node, and stress assignments. In this case, the reference point is taken to be $U$ at $(r,\varphi)=(r_0,0)$, and the geodesics are constructed in directions forming angles $\eta$ and $2\eta$ below the rod $LR$.
We perform a numerical evaluation for this configuration as well, adopting $r_s=0.1$ and $(r_0,\varphi_0,\eta)=(0.3,45^\circ,15^\circ)$. In this case, we obtain
\begin{equation}
\frac{F_r}{S} = 0.032.
\label{eq:inverted_triangle_numeric_force}
\end{equation}
Thus, an upward effective force also arises in the inverted triangle configuration. The configuration shown in Fig.~\ref{fig:inverted-triangle-configuration} is drawn based on this numerical result.

\subsection{Perturbative analysis of the triangle and inverted triangle configurations}

We perform a perturbative analysis for both the triangle and inverted triangle configurations in the same regime as for the diamond configuration, namely $r_0 \gg r_s$ and $\varphi_0 \ll 1$. In both cases, the leading nonvanishing term in the expansion of the radial force $F_r$ appears at order $r_s \varphi_0^3$, yielding
\begin{equation}
\frac{F_r}{S} = r_s\left[\frac{1}{4r_0} \tan(\eta)\tan(2\eta)\,\varphi_0^3 + O(\varphi_0^4)\right] + O(r_s^2).
\end{equation}
Thus, for both the triangle and inverted triangle configurations, the first nonzero contribution to the radial force arises at order $r_s \varphi_0^3$, in agreement with the result for the diamond configuration given in Eq.~\eqref{eq:diamond_force_pert}.

\section{Discussion}

In this work, we have considered static structures composed of geodesic rods in Schwarzschild spacetime and have shown that an effective force arises due to the incomplete cancellation of internal stresses. In order to satisfy force balance at each vertex, a non-gravitational downward external force must be applied to the structure, indicating that an effective force acting against gravity---namely, a buoyancy-like force---emerges at the level of the entire structure. A key feature of this effect is that it arises in a static configuration without any time-dependent internal motion.
We note that the direction of the resulting force depends on the choice of internal stresses on the geodesic rods: exchanging tensile and compressive stresses reverses the sign of the force.

Such a phenomenon can be understood within the existing framework of extended-body dynamics in curved spacetime. The curved-spacetime swimming effect proposed by Wisdom refers to the displacement of the center of mass induced by cyclic internal deformations in the absence of external forces. In contrast, the effect considered here does not involve time dependence and instead arises purely from the coupling between internal stresses and the gravitational field. In the multipole formalism developed by Dixon, it is known that the coupling between internal structure and spacetime curvature contributes to the equations of motion, and the present result can be regarded as a concrete realization of this mechanism. Furthermore, as pointed out by Israel~\cite{Israel1966}, stresses can make nontrivial contributions to force balance in a gravitational field. However, the essential feature of the effect discussed here lies not in the magnitude of the stress itself, but in the geometrically induced asymmetry in the directions of stresses within the structure.

On the other hand, the buoyancy-like effect obtained in this study does not provide a mechanism to support a structure against gravity. Since the result takes the form $F_r \propto S$, it may appear that the buoyancy can be enhanced arbitrarily by increasing the internal stress $S$. However, an increase in stress necessarily leads to an increase in the gravitational source through the energy--momentum tensor, and the self-weight of the structure increases accordingly. Therefore, the upward effective force cannot compensate for this increase. Moreover, physically reasonable matter must satisfy the dominant energy condition (DEC), which constrains the stress by the energy density ($|p| \leq \rho$). As a result, it is not possible in principle to enhance the buoyancy arbitrarily by increasing the internal stress. Separately, since $F_r/S$ scales as $(r_s/r_0)\varphi_0^3$, the magnitude of the effect is extremely small under realistic conditions, and it is therefore practically impossible to detect experimentally.

In summary, although the effect is extremely small under realistic conditions and practically undetectable, it demonstrates that internal stresses can manifest as an effective force through geometrical coupling in a spacetime with a curvature gradient, thereby revealing a novel aspect of the interplay between internal structure and spacetime curvature in extended-body dynamics.

\section{Conclusion}

In this work, we have considered static structures composed of geodesic rods in Schwarzschild spacetime and have shown that an effective force opposing gravity can arise from the coupling between internal stresses and the gravitational field. A key feature of this effect is that it does not require any time-dependent internal motion, but instead arises solely from the geometry of the structure and the distribution of internal stresses.

On the other hand, this buoyancy-like effect does not provide a mechanism to support a structure against gravity. An increase in internal stress necessarily leads to an increase in both the gravitational source, through the energy--momentum tensor, and the self-weight of the structure. In addition, physical constraints such as the energy conditions limit the magnitude of the stress, preventing any arbitrary enhancement of the effect. Furthermore, since its magnitude is suppressed by the scaling $(r_s/r_0)\varphi_0^3$, the effect is extremely small under realistic conditions and is practically impossible to detect experimentally.

The present result provides a simple and explicit example demonstrating that, in curved spacetime, the coupling between internal stresses and spacetime curvature can manifest as an effective force in the dynamics of extended bodies.

\appendix
\renewcommand{\theequation}{\thesection.\arabic{equation}}
\renewcommand{\theHequation}{\thesection.\arabic{equation}}
\setcounter{equation}{0}
\section{Stress propagation in a geodesic rod}
\label{app:stress-propagation}

\subsection{Stress propagation without self-weight}
\label{appsec:stress-without-self-weight}

In this appendix, we derive the propagation of stress in a geodesic rod whose self-weight can be neglected. The rod is modeled as a one-dimensional elastic body that supports stress only along its axis, with sufficiently large bending rigidity so that the rod axis remains straight. The rod is assumed to be part of a larger structure: external forces may act at its endpoints, while no external force density acts in the interior.

On the equatorial plane ($\theta=\pi/2$) of Schwarzschild spacetime, the metric can be written in the form
\begin{equation}
ds^2=-\alpha^2dt^2+\gamma_{ij}dx^i dx^j,
\qquad
\gamma_{ij}dx^i dx^j=\alpha^{-2}dr^2+r^2d\phi^2,
\label{eq:app_metric}
\end{equation}
where $\alpha=\sqrt{1-r_s/r}$, and $\gamma_{ij}$ denotes the spatial metric. Let $\lambda$ be the spatial arc length along the rod axis, defined by
\begin{equation}
d\lambda^2=\gamma_{ij}dx^i dx^j.
\label{eq:app_arclength}
\end{equation}
Let $t^i$ be the unit tangent vector along the rod axis, satisfying $\gamma_{ij}t^i t^j=1$. We define the corresponding spacetime vector as
\begin{equation}
\tau^\mu=(0,t^i),
\label{eq:app_tau}
\end{equation}
which represents a purely spatial direction and is orthogonal to the four-velocity of a locally static observer, $u^\mu=(\alpha^{-1},0,0,0)$, satisfying $\tau_\mu\tau^\mu=1$.

Let $T(\lambda)$ denote the stress measured by a locally static observer along the rod axis. Assuming that the rod supports stress only along its axis, the energy--momentum tensor is then given by
\begin{equation}
T^{\mu\nu}=-T\tau^\mu\tau^\nu,
\label{eq:app_emtensor}
\end{equation}
where $T>0$ corresponds to tension and $T<0$ to compression.

From the conservation law $\nabla_\mu T^{\mu\nu}=0$, we obtain
\begin{equation}
0=\nabla_\mu(T\tau^\mu\tau^\nu)=(\nabla_\mu T)\tau^\mu\tau^\nu+T(\nabla_\mu\tau^\mu)\tau^\nu+T\tau^\mu\nabla_\mu\tau^\nu.
\label{eq:app_cons}
\end{equation}
Projecting onto the spatial components orthogonal to $u^\mu$, this reduces to
\begin{equation}
\left(\frac{dT}{d\lambda}+T\nabla_\mu\tau^\mu\right)\tau^a+T(\nabla_t\tau)^a=0,
\label{eq:app_projected}
\end{equation}
where
\begin{equation}
\frac{dT}{d\lambda}=\tau^\mu\nabla_\mu T,
\qquad
(\nabla_t\tau)^a=\tau^\mu\nabla_\mu\tau^a.
\label{eq:app_defs}
\end{equation}
In Schwarzschild spacetime, the relation $\Gamma^0{}_{0i}=\partial_i\ln\alpha$ gives
\begin{equation}
\nabla_\mu\tau^\mu=t^i\partial_i\ln\alpha=\frac{d\ln\alpha}{d\lambda},
\label{eq:app_divtau}
\end{equation}
where, under the assumptions of a uniform one-dimensional rod, the spatial divergence $D_i t^i$ vanishes; here $D_i$ denotes the covariant derivative associated with the spatial metric $\gamma_{ij}$.

Since $\tau_a(\nabla_t\tau)^a=\frac{1}{2}\nabla_t(\tau_a\tau^a)=0$, the vector $(\nabla_t\tau)^a$ is orthogonal to $\tau^a$. Therefore, the tangential and normal components of Eq.~\eqref{eq:app_projected} must vanish independently, yielding
\begin{equation}
\frac{dT}{d\lambda}+T\frac{d\ln\alpha}{d\lambda}=0,
\qquad
(\nabla_t\tau)^a=0.
\label{eq:app_split}
\end{equation}
The first equation in Eq.~\eqref{eq:app_split} gives
\begin{equation}
\frac{d}{d\lambda}(\alpha T)=0,
\label{eq:app_alphaT_const1}
\end{equation}
and hence
\begin{equation}
\alpha T=\mathrm{const.}
\label{eq:app_alphaT_const2}
\end{equation}
Thus, the stress measured by a locally static observer varies as $T\propto\alpha^{-1}$ along the rod, while $\alpha T$ is conserved.

The second equation in Eq.~\eqref{eq:app_split} implies that the spatial curve $x^i(\lambda)$ representing the rod axis is a geodesic with respect to the spatial metric $\gamma_{ij}$. Indeed, using $t^i=dx^i/d\lambda$, the condition $t^jD_j t^i=0$ yields
\begin{equation}
\frac{d^2x^i}{d\lambda^2}+{}^{(3)}\Gamma^i{}_{jk}\frac{dx^j}{d\lambda}\frac{dx^k}{d\lambda}=0.
\label{eq:app_spatial_geodesic}
\end{equation}
On the equatorial plane, with $x^i=(r,\varphi)$, this coincides with the spatial geodesic equation used in the main text.

Therefore, for a geodesic rod without self-weight, the stress propagates according to Eq.~\eqref{eq:app_alphaT_const2}, and the rod axis follows a spatial geodesic.
Here, $\alpha T$ corresponds to the stress $T$ measured locally, converted by the lapse factor to the normalization defined with respect to infinity. Therefore, when expressed in this quantity, the forces acting at the two endpoints of the rod are equal in magnitude and opposite in direction along the rod axis.
In the main text, the tensions and compressive forces in the structure are treated in terms of this quantity defined with respect to infinity. Since force balance at each vertex is evaluated locally, the lapse factor cancels as a common factor and need not be considered explicitly.

\subsection{Stress propagation with self-weight}

We now consider the case where the mass of the rod is not neglected. The rod has a linear mass density $\rho$ and a constant cross-sectional area $\sigma$. The energy density measured by a locally static observer is then given by $\varepsilon=\rho/\sigma$.
We assume that the rod is uniform, so that $\rho$ and $\sigma$, and hence $\varepsilon$, are constant along the rod.

Among the components of the gravitational acceleration acting on the rod, those perpendicular to the rod axis are assumed to be supported by an external guiding mechanism (hereafter, the guide), which does not exert any force along the rod axis. The mass of the guide itself is neglected.
The energy--momentum tensor is then given by
\begin{equation}
T^{\mu\nu}=\varepsilon u^\mu u^\nu-T\tau^\mu\tau^\nu,
\label{eq:app_self_emtensor}
\end{equation}
and the conservation law, including the external force density $f_{\mathrm{ext}}^\nu$, is written as
\begin{equation}
\nabla_\mu T^{\mu\nu}=f_{\mathrm{ext}}^\nu.
\label{eq:app_self_cons}
\end{equation}
Projecting onto the spatial components orthogonal to $u^\mu$, we obtain
\begin{equation}
\left(\frac{dT}{d\lambda}+T\frac{d\ln\alpha}{d\lambda}\right)\tau^a+T(\nabla_t\tau)^a=f_g^a-f_{\mathrm{ext}}^a,
\label{eq:app_self_projected}
\end{equation}
where we have used $\nabla_\mu\tau^\mu=d\ln\alpha/d\lambda$, as in Sec.~\ref{appsec:stress-without-self-weight}, following the same assumption of a uniform one-dimensional rod, and $f_g^a\equiv\varepsilon u^\mu\nabla_\mu u^a$ represents the gravitational force density acting on the rod, as measured by a locally static observer.
Defining $a_i\equiv D_i\ln\alpha$ and $a^i=\gamma^{ij}a_j$, we have
\begin{equation}
f_g^i=\varepsilon a^i.
\label{eq:app_self_fg}
\end{equation}
Since the guide provides only normal forces, we have $t_i f_{\mathrm{ext}}^i=0$. In the normal direction, the rod is supported so that $f_{\mathrm{ext}\perp}^i=f_{g\perp}^i$.
The normal component then yields
\begin{equation}
(\nabla_t\tau)^a=0,
\label{eq:app_self_geodesic}
\end{equation}
so that the rod axis remains a spatial geodesic even when self-weight is included.
On the other hand, along the tangential direction, $f_{\mathrm{ext}\parallel}=0$, and Eq.~\eqref{eq:app_self_projected} gives
\begin{equation}
\frac{dT}{d\lambda}+T\frac{d\ln\alpha}{d\lambda}=t_a f_g^a.
\label{eq:app_self_tangent}
\end{equation}
Using $f_g^a=\varepsilon a^a$ with $a^a=D^a\ln\alpha$, we obtain
\begin{equation}
t_a f_g^a=\varepsilon\frac{d\ln\alpha}{d\lambda}.
\label{eq:app_self_tfg}
\end{equation}
Therefore, Eq.~\eqref{eq:app_self_tangent} becomes
\begin{equation}
\frac{dT}{d\lambda}+T\frac{d\ln\alpha}{d\lambda}=\varepsilon\frac{d\ln\alpha}{d\lambda}.
\label{eq:app_self_eqn}
\end{equation}
Since $\varepsilon=\mathrm{const.}$, this yields
\begin{equation}
\alpha(T-\varepsilon)=\mathrm{const.}
\label{eq:app_self_const}
\end{equation}
Thus, when self-weight is included, the rod axis still follows a spatial geodesic, while the conserved quantity along the rod is modified from $\alpha T$ to $\alpha(T-\varepsilon)$.

This result shows that the effect of the rod's self-weight can be treated independently in the analysis of stress imbalance. In other words, the propagation law is obtained by replacing $T$ with $T-\varepsilon$, while the difference in the stress defined with respect to infinity at the two endpoints $A$ and $B$, $\alpha_A T_A-\alpha_B T_B$, corresponds to the integral of the tangential component of the gravitational force along the rod. Together with the supporting force from the guide, this accounts for the gravitational force acting on the rod.
Therefore, the analysis including self-weight can be treated equivalently as that for a weightless rod, with the gravitational force acting on the rod added independently.

\section*{Acknowledgements}

The author acknowledges the use of ROOT, developed at CERN, for numerical calculations and figure generation. Analytical and perturbative calculations were performed using Mathematica.

\end{document}